# My Boss the Computer: A Bayesian analysis of socio-demographic and cross-cultural determinants of attitude toward the Non-Human Resource Management

Mantello Peter, Manh-Tung Ho, Minh-Hoang Nguyen, & Quan-Hoang Vuong

*Un-peer-reviewed manuscript version 1.1*

January 21, 2021

*Beppu, Oita, Japan*

# My Boss the Computer: A Bayesian analysis of socio-demographic and cross-cultural determinants of attitude toward the Non-Human Resource Management

**Mantello Peter[1], Manh-Tung Ho[1,2,3*], Minh-Hoang Nguyen [1,2], Quan-Hoang Vuong [2,4*]**

[1] Ritsumeikan Asia Pacific University, Beppu, Oita, Japan, 874-8577

[2] Centre for Interdisciplinary Social Research, Phenikaa University, Yen Nghia Ward, Ha Dong District, Hanoi 100803, Vietnam

[3] Institute of Philosophy, Vietnam Academy of Social Sciences, 59 Lang Ha Street, Ba Dinh District, Hanoi 100000, Vietnam

[4] AI for Social Data Lab, Vuong & Associates, 3/161 Thinh Quang, Dong Da District, Hanoi 100000, Vietnam

Corresponding authors: Manh-Tung Ho (tung.homanh@phenikaa-uni.edu.vn), Quan-Hoang Vuong (hoang.vuongquan@phenikaa-uni.edu.vn)

**Acknowledgments**

This study is part of the project "*Emotional AI in Cities: Cross Cultural Lessons from UK and Japan on Designing for an Ethical Life*" funded by JST-UKRI Joint Call on Artificial Intelligence and Society (2019). The authors would like to thank all the APU faculty members that helped us distribute the survey.

**Abstract**

Human resource management technologies have moved from biometric surveillance to emotional artificial intelligence (AI) that monitor employees' engagement and productivity, analyze video interviews and CVs of job applicants. The rise of the US$20 billion emotional AI industry will transform the future workplace. Yet, besides no international consensus on the principles or standards for such technologies, there is a lack of cross-cultural research on future job seekers' attitude toward such use of AI technologies. This study collects a cross-sectional dataset of 1,015 survey responses of international students from 48 countries and 8 regions worldwide. A majority of the respondents (52%) are concerned about being managed by AI. Following the *hypothetico-deductivist* philosophy of science, we use the MCMC Hamiltonian approach and conduct a detailed comparison of 10 Bayesian network models with the PSIS-LOO method. We consistently find having a higher income, being male, majoring in business, and/or self-rated familiarity with AI correlate with a more positive view of emotional AI in the workplace. There is also a stark cross-cultural and cross-regional difference. Our analysis shows people from economically less developed regions (Africa, Oceania, Central Asia) tend to exhibit less concern for AI managers. And for East Asian countries, 64% of the Japanese, 56% of the South Korean, and 42% of the Chinese professed the trusting attitude. In contrast, an overwhelming majority of 75% of the European and Northern American possesses the worrying/neutral attitude toward being managed by AI. Regarding religion, Muslim students correlate with the most concern toward emotional AI in the workplace (***β_Islam_Attitude***'s mean =-0.16, sd =0.10; ***β_Buddhism_Attitude***'s mean =-0.05, sd =0.07; ***β_Christian_Attitude*** 's mean = -0.10, sd= 0.09). When religiosity is higher, the correlation becomes stronger for Muslim and Buddhist

students. This paper adds a cross-cultural perspective to the literature, which is currently skewed toward country-specific and profession-specific samples.

## Introduction

Emotion detection or emotion-sensing artificial intelligence (AI) is a fast-growing industry, which is reportedly worth around a US$20 billion industry (Telford, 2019). This emerging technology is already transforming how we live and work, according to Nature Index (Condie & Dayton, 2020). Given the on-going complexities and pressures of the COVID-19 pandemic, there is growing interest in surveillance technologies both in the public sphere and workplace (Roussi, 2020). This article examines various aspects of non-human resource management using emotional AI technologies. It then presents a Bayesian analysis of future job seekers' cross-cultural perceptions of AI recruitment, work monitoring, and possible threats to the employee's autonomy.

Workplace surveillance technologies have a long history, as Ball (2010) stated, "organization and surveillance go hand in hand," and "workplace surveillance can take social or technological forms" (Ball, 2010, p. 87). Indeed, early scholarship on workplace surveillance considers particular arrangements of the workplace that allow one manager to watch over a large number of his or her employees, i.e., the panopticon model (Galič et al., 2017). Frederick Winslow Taylor's seminal observational studies allowed managers to determine the most efficient way workers could do their jobs (Kiechel, 2012). By the late 20th century, the task of human resource management to ensure productivity in the workplace was augmented through technological developments in biometric surveillance (i.e., fingerprints, gaits, etc.) to identify and monitor individuals. Today, biometric surveillance has become the norm in many modern workplaces. According to Marciano (2019), biometric surveillance prioritizes the body over the mind as it treats the reading of bodily information as a more reliable and "objective."

However, recent advances in affective computing, with the promise to accurately detect human emotions, put the mind back at the ontological center of workplace surveillance. The term *'affective computing'* was coined by Rosalin Picard (1995) to describe her development of computational technologies that sense, gauge, and respond to human mental states, including emotions and dispositions. With advances in affective computing, especially when combined with AI and machine learning, affective computing is now reaching a point where it is capable of doing what was once only thought possible in humans, which is to read the emotional state of someone. Simply put, emotional AI is the ability of machines and devices to extract data of a person's emotional state by reading their facial expressions, body language, skin conductance level, eye movement, voice tone, respiration, and heart rate variability, as well as machine learning of images and words (Larradet et al., 2020; McStay, 2018; Richardson, 2020; Rukavina et al., 2016). Emotional AI has a wide range of applications. For example, Grammarly's algorithm can detect tone via texts. Spotify can suggest playlists by sensing a person's mood, or Amazon's home assistant, Alexa, can detect through voice analytics the emotional state of its users (Richardson, 2020). Honda and Softbank, as a response to the spike in accidents by elderly drivers in Japan, have co-created the 'Emotion Engine', which detects if a driver is drowsy, distracted, or stressed, to increase driver safety. The Japanese company 'Empath' has developed voice recognition software that reads both its employees' and customers' moods to increase workplace efficiency and monitor workers' well-being. NEC has developed software for McDonalds to gauge customer sentiment while they are looking at digital menus to optimize the customer's experience while at the same time increasing sales.

Moreover, emotional AI is increasingly being marketed as a surveillance tool for policing and border security. Yet, like many "surveillance tools" that rely on facial recognition, Amazon has stopped providing its REKOGNITION technology to the government, but it is still being sold to commercial customers (Chouinard et al., 2019). Likewise, emotional AI technologies have been used to exploit the emotionality of political communication. A good example is Cambridge Analytica, which used micro-targeted ads to exploit and manipulate voter emotions in Brexit and US elections three years ago (Wylie, 2019).

The use of emotional AI in workplace settings is growing worldwide. For example, about 25% of the top 131 companies in South Korea have already used or planned to use AI in their hiring process (Partner, 2020). Large corporations such as IBM, Unilever, and Softbank are already using AI to recruit and analyze which behaviors or traits expressed by the interviewees could predict future good performers (Richardson, 2020). Critically, beyond the recruitment process, emotional AI technologies monitor stress, alertness, and the level of engagement and attention of workers (Larradet et al., 2020; Suni Lopez et al., 2019).

While emotional AI technologies are becoming more pervasive in society, their impact on individuals in the workplace is understudied. Top corporates that sell emotional AI products such as Affectiva or Empath claim an accuracy of $90^{th}$ percentile in reading correct emotions from facial expressions and other physiological signals (Heaven, 2020). Both large and small companies are joining the race to develop and sell these AI products, driving the industry's value to USD 20 billion, reportedly (Telford, 2019). Yet besides there being no international agreement on the standards and principles for emotional AI technologies, the datasets on emotion, used to train the algorithms, are being coded and grown by humans using crowdsourcing platforms under Paul Ekman (1999)'s now contentious *eight basic universal emotions* (i.e., anger, fear, sadness, disgust, surprise, anticipation, trust, and joy) (Mitchell, 2019; Mohammad & Turney, 2013; Yue et al., 2019). This leads to two glaring concerns over the accuracy and cultural bias inherent in the universal application of emotional AI.

McStay (2018, p. 4) coined the term *"machinic verisimilitude"* to express sympathy for the technology and business communities, which have yet to deal properly with the social constructionist complexities of ethnocentric, context-dependent views of emotions. In stark contrast, in a recent article in *Nature Neuroscience,* seven top-tiered researchers proposed several different approaches to define and investigate fear, a seemingly simple and universal emotion (Mobbs et al., 2019). Recent empirical and theoretical developments in the studies of emotion have questioned the validity of the so-called *universality thesis of emotion* (see Gendron et al. (2018)) as the modus operandi for the *emphatic media* industry.

Researchers such as Lisa Barrett and Hoemann et al.'s work on *'constructed emotion'* show that the communication and inference of anger, fear, disgust, or any other Ekman's basic emotions have significant cultural and contextual variations as a result of reviewing more than 1,000 academic articles on emotional expression (Barrett et al., 2019). Moreover, modes of emoting evolve since cultures are dynamic and unbounded, with the constant cultural transmission, learning, and unlearning (Boyd et al., 2011; Henrich, 2020; Vuong et al., 2020; Vuong, 2016). This truism about culture challenges the traditional/normative and static ways of structuring emotion datasets into Ekman's eight basic types, the *valence* dimension (i.e., positive, neutral, negative sentiments), the *arousal* dimension (i.e., bored versus excited), favored by the tech companies (McStay, 2018). Unlike how current machines read emotions, a vast body of literature shows the human reading of emotions and their intensity depends on various external factors such as body movement, tone of voices, tones of skin, or the background scene (Benitez-Quiroz et al., 2018; Chen et al., 2018) as well as personal experiences and cultural setting (Barrett, 2017). More importantly, the fact that many job seekers are now aware of AI-hiring and starting to game the algorithms by presenting themselves differently using different words than they naturally would (Borsellino, 2020) makes the concern over accuracy even graver. This can be witnessed by the plethora of amateur videos on YouTube that teach users 'how to beat AI recruiting'.

Accuracy concern aside, the human-coded crowdsourcing datasets for training emotional AI raise serious concerns and algorithmic bias issues. For example, Rhue (2019) shows two facial recognition algorithms, *Microsoft AI* and *Face++*; both have a systematic bias about reading emotions such as anger and contempt, especially when interpreting emotions of different races. Purdy *et al.* (2019) explain the potential for algorithmic bias arising from simple facts; for instance, 89% of civil engineers and 81% of first-line police are male in America. Thus, an algorithm trained on these professions' datasets will struggle to read female employees' emotions in the field (Purdy et al., 2019). Timnit Gebru, a former AI ethics researcher, was fired from Google over a co-authored paper she wrote concerning the risks of training a large language model (Singh, 2020). One of Gebru and her colleagues' main discoveries was that facial recognition systems are less accurate at identifying women of color (Buolamwini & Gebru, 2018).

The lack of accuracy and embedded bias in current algorithmic decision-making suggests the need for greater accountability of AI technology and a systematic way of understanding how people perceive the introduction of the so-called "AI managers." Hence, in this paper, we survey a large body of students, 1015 future job seekers, from 48 countries and 11 regions around the world to understand how various socio-demographic, cultural, and economic factors influence perception and attitude toward three aspects of the AI managers: job entry gatekeeping, work monitoring, and the autonomy threat.

**Literature Review**

There is a lack of coherent, empirical cross-cultural studies literature on perception toward AI. Critically, what little that exists focuses on AI applications' perception as the main tool to optimize non-human aspects of productivity in the modern workplace. This section examines the current findings of AI-perception literature.

*Students' and professional's attitude toward AI*

Of the few studies on the perception of AI in the modern workplace, it is clear the research methods to measure awareness of AI, and its effects are still in an early stage of development. Brougham and Haar (2017) designed a measurement instrument called STARA awareness scale, which stands for Smart Technology, Artificial Intelligence, Robotics, and Algorithms (STARA), to measure the employees' perceptions of the future workplace. Testing the STARA scale on 120 employees in New Zealand, Brougham and Haar (2017) found the more employees were aware of STARA, and what it was meant for their job, the lower organizational commitment and career satisfaction the authors found among them. These findings are concurrent with previous studies that have examined the relationship between biometric surveillance and employee trust in the workplace (Ball, 2010; Marciano, 2019).

One consistent finding in the literature is that people have little concern over job loss due to AI (Brougham & Haar, 2017; Pinto dos Santos et al., 2019; Sarwar et al., 2019; Shen et al., 2020). Indeed one particular survey of 487 pathologists shows that nearly 75% of the respondents show excitement and interest in the prospect of AI integration in their work (Sarwar et al., 2019). Similarly, Pinto dos Santos et al. (2019) surveyed 263 medical students and found although most respondents thought AI could revolutionize (77%) and improve radiology (86%), 83% did not think human physicians would be replaced. Likewise, Shen et al. (2020) found nearly 96% conducts 1,228 Chinese dermatologists from 30 provinces and regions in a web-based survey that believed AI roles are to assist with diagnosis and treatment. On the other hand, while a survey of 484 responses of 19 UK universities medical students show that a majority of students (88%) will play an important role in healthcare yet, nearly half the respondents express their concern about job loss in radiology due to AI (Sit et al., 2020). However, overall, most medical students' population found students express interest and desire to receive more AI training in their degree (Bin Dahmash et al., 2020; Pinto dos Santos et al., 2019; Sit et al., 2020).

A mixed-method study on the perception of the proliferation of AI chatbots in healthcare showed that most internet users are receptive to this new technology; however, there was a level of hesitancy due to concerns about cyber-security (hacking/theft of personal data), privacy (third party sharing of information), accuracy, and questions concerning the ability for empathy of AI (Nadarzynski et al., 2019). In the same vein, Gao et al. (2020) found an underlying reason for the negative public perception of medical AI was a fundamental distrust in profit-orientated companies and nascent stage of the technology as the result of tracking 2,315 social media posts in the Chinese platform Sina Weibo related to AI in 2017. They also found that of 965 posts where attitudes were expressed, nearly 60% were positive, and 6.2% were negative about using AI in medical settings.

*Public perception toward AI*

Public perception toward AI has been shaped by multiple factors such as socio-cultural backgrounds, fictional representation in popular culture, public positions of the experts in the field, political events and scandals, perception of economic and political rivals. Historically speaking, public perception of AI has been greatly influenced by fictional representation in popular novels, film, and television, many of which represent AI as an innately dystopian form of synthetic intelligence such as *Colossus: The Forbin Project* (1970), *War Games* (1983), *Terminator* (1984), *I, Robot* (2004), *and Ex Machina* (2014). On the other hand, Asian people tend to associate AI in a more favorable capacity, such as *Mighty Atom (Astro Boy)* and the beloved manga/animation character *Doraemon* (Robertson, 2017). However, there is contradictory evidence from the empirical literature. For instance, Bartneck *et al.*'s (2007) study found that among nearly 500 participants, people from the US are the most positive toward robots, while Japanese and Mexican are more negative toward robots.

Mass media can also shape public perception of AI. Neri and Cozman (2020) have shown that cautious opinions offered by technology pundits such as Stephen Hawking or Elon Musk could change AI risk evaluation. Moreover, a study of media discussions on AI in the New York Times taken over 30 years showed a progressive increase in concern for the human loss of control over AI, ethical concerns about the role of AI in society, and displacement of human activity in the workforce (Fast & Horvitz, 2017). On the other hand, the optimistic view of AI application in healthcare and education was also found to increase. Using the NexisUni database, Ouchchy et al. (2020) found the tone of media coverage of AI's ethical issues were initially optimistic and enthusiastic in 2014 and became more critical and balanced until 2018, with the privacy issue being the most salient aspects of this debate.

Controversial political events such as the Cambridge Analytica case or the yet to be passed Algorithmic Accountability Act in the US can also shape the public discourse of the risk of AI misuse. The UK-based data broker company fell into disrupting when the public was made aware that its various parent companies, such as SCL Elections Ltd., had executed psychological operations (psy-ops), powered by harvesting massive social media data with algorithms to micro-target and change individual behaviors, in more than 200 elections around the world, mostly in underdeveloped countries (Kaiser, 2019; Wylie, 2019). Bakir (2020) assessed the profiling offered by the company to the Leave.EU campaign in the 2016 Brexit Referendum and showed such practice has both *deceptive* and *coercive* features. Because of public backlash, Cambridge Analytica is now a nonoperational political data analytics company.

Since then, in surveys around the world, where people are aware of digital micro-targeting practices, they have expressed a clear desire for action against technologies that exploit the emotionality of voters in political campaigns (Woolley & Howard, 2018). Yet, many still do not realize that behind the political adverts they see online, AI-powered micro-targeting tools are deployed (Bakir, 2020). For example, according to a YouGov survey in 2019, while 58% of the UK national sample were against tailoring political adverts, 31% of the UK sample were unaware of these problems (ORG, 2020). In response to the growing public concerns over the manipulativeness and intrusiveness of the AI-powered digital political and marketing campaign, politicians in advanced democracies have started to push for legislation that increases companies' transparency and accountability to build and deploy these AI systems (Badawy et al., 2018). Legislations such as the EU Digital Services Act, or the Algorithmic Accountability Act and the Filter Bubble Transparency Act in the US, or the German *Medienstaatsvertrag* (State Media Treaty) have sparked heated public debates and received supports from certain political factions and stakeholder groups (Rieder & Hofmann, 2020). Moreover, crucial data are absent on public debates about AI governance in developing countries.

As such, the current literature on the perception of AI shows three major areas of concern. First, there is a clear lack of cross-cultural and cross-regional comparison. Second, empirical studies on the subject indicate a shortage of consistent measuring and testing instruments for AI perception determinants. Finally, the absence of studies on the impact of emotion-sensing technologies in the workplace suggests a strong necessity for further research to fill the existing intellectual vacuum. These are the three areas where this study seeks to contribute.

### Research questions

Based on the literature, we ask the following two major research questions:

RQ1: How do socio-demographic factors influence self-rated familiarity with AI?

RQ2: How do socio-demographic factors (sex, income, religion, religiosity, major, school year, regions) and self-rated knowledge influence respondents' perception toward the AI managers?

## Methodology

### Methods

#### Data collection

The survey was distributed in 14 online classes in Ritsumeikan Asia Pacific University (APU), Beppu, Oita, Japan, and a public Facebook group of APU students. APU is the most international campus in Japan, with students coming from 91 countries and regions in the world as of the academic year 2019. The first round of distribution is from July 15th to August 5th, 2020, and the second round from October 10th to December 10th, 2020. The research assistant came into online zoom classes and explained the purpose of the study is to measure the broad statistical patterns of how students' attitudes and knowledge about emotional AI vary according to their socio-demographic factors. The students were also told that the survey's participation would be voluntary, and all responses would be anonymized.

*Data treatment*

**Table 1:** Explanation of the data treatment procedure

| Variables | Variable type | Summary | Remarks/Survey questions |
|---|---|---|---|
| **Outcome variable** | | | |
| *Attitude* | Continuous | The attitude toward the application of emotional AI in the workplace ("1" strongly disagree/very worried, "5" strongly agree/not worried). | The *Attitude* variable is first calculated by averaging the answers of **three** Likert-scale questions, 1) Do you agree that a company manager should use AI/smart algorithms to measure employees' performances? 2) Do you agree that a company manager should use AI/smart algorithms to screen job applicants? 3) Are you worried about protecting your autonomy at work due to the wider application of AI/smart algorithms? |
| *Familiarity* | Continuous | Taking the average of the four questions on the right side ("1" being Not familiar; "5" being Very familiar) | The variable attitude is calculated by averaging the answers of **four** Likert-scale questions: 1) How familiar are you with coding/programming? 2) How familiar are you with the topic of emotional AI? |

| | | | 3) How familiar are you with the concept of smart cities? |
| | | | 4) How familiar are you with the topic of Artificial Intelligence (AI)? |
| **Predictive variable** | | | |
| *SchoolYear* | Ordinal/ Continuous | $1^{st}$, $2^{nd}$, $3^{rd}$, and $4^{th}$ year | |
| *Sex* | Binary | Male ("1") vs. Female ("0") | Respondents choose their biological sex. |
| *Income* | Ordinal/ Continuous | low ("1"), middle ("2"), and high ("3") | Self-perceived level of household income. |
| *Major* | Binary | Social studies ("0") vs. Business ("1") | Students are asked to specify their majors. |
| *Religions* | Binary | Christianity:"1"if identified<br>Islam: "1" if identified<br>Buddhism: "1" if identified | Respondents are asked to specify their official religion and the lack thereof. There are very few Jewish and Shinto-ist respondents; thus they are not included in our analyses. |
| *Religiosity* | Binary | "1" for the very religious, "0" for the non-religious or mildly religious. | Respondents are asked to choose their level of religiosity. |

*Bayesian multi-level analysis*

***Model construction***

Following the recent guidelines on conducting Bayesian inference, ten models are first constructed by gradually adding more variables and levels (Aczel et al., 2020; Gelman & Shalizi, 2013; Vuong et al., 2018; Vuong et al., 2020). Then, the models are fitted with the data using the Bayesian Hamiltonian Monte Carlo approach. The Bayesian priors of all parameters' distribution are set as default, which is '*uninformative*' (Andrieu et al., 2003; McElreath, 2020). The equations of the models are presented in Table 2 below:

**Table 2:** Equations of the models

| Model | Equation |
|---|---|
| 1 | $Attitude \sim Income + SchoolYear + Sex + Major$ |
| 2 | $Attitude \sim Christianity + Islam + Buddhism$ |
| 3 | $Attitude \sim Christianity + Islam + Buddhism + Christianity\_Religiosity + Islam\_Religiosity$ $+ Buddhism\_Religiosity$ |
| 4 | $Attitude \sim Income + SchoolYear + Sex + Major + Christianity + Islam + Buddhism$ $+ Christianity\_Religiosity + Islam\_Religiosity + Buddhism\_Religiosity$ |
| 5 | $Familiarity \sim Income + SchoolYear + Sex + Major$ |
| 6 | $Familiarity \sim Christianity + Islam + Buddhism$ |
| 7 | $Familiarity \sim Christianity + Islam + Buddhism + Christianity\_Religiosity + Islam\_Religiosity$ $+ Buddhism\_Religiosity$ |
| 8 | $Familiarity \sim Income + SchoolYear + Sex + Major + Christianity + Islam + Buddhism$ $+ Christianity\_Religiosity + Islam\_Religiosity + Buddhism\_Religiosity$ |
| 9 | $Attitude \sim Familiarity + Income + SchoolYear + Sex + Major + Christianity + Islam + Buddhism$ $+ Christianity\_Religiosity + Islam\_Religiosity + Buddhism\_Religiosity$ |
| 10 | $Attitude \sim \text{alpha}[Region_{varint}] + Familiarity + Income + SchoolYear + Sex + Major + Christianity$ $+ Islam + Buddhism + Christianity\_Religiosity + Islam\_Religiosity$ $+ Buddhism\_Religiosity$ |

As can be seen, the models are gradually expanded, tested, and compared, which are aligned with the *hypothetico-deductivist* philosophy of science championed by Gelman & Shalizi (2013). It is worth noted that models with both religion and religiosity variables are nonlinear to avoid confounding effects. Model 10 is then a multi-level model; thus, it is the most complex model with the *Region* variable functions as the varying-intercept. There are all other variables present in this model. Multi-level modeling has a natural fit with Bayesian analysis as it assigns probability distributions to the varying regression coefficients (Spiegelhalter, 2019).

Moreover, the multi-level fitting model also helps improve the estimate for imbalance in sampling and explicitly study the variation among groups. Partial pooling (or adaptive pooling) is another advantage of multi-level modeling. This kind of pooling enables us to produce less underfit estimates than complete pooling and overfit than no-pooling (Gelman & Hill, 2006; McElreath, 2020).

Finally, to guard against overfitting and find the model best fitted with the data, the models are compared in detail using the Pareto smoothed importance-sampling leave-one-out cross-validation (PSIS-LOO) approach (Vehtari et al., 2017). We will compare the models' weights by computing the Pseudo-BMA weights without Bayesian bootstrap, Pseudo-BMA+ weights with Bayesian bootstrap, and Bayesian stacking weights (Vehtari & Gabry, 2019; Yao et al., 2018).

**Results**

*Descriptive statistics*

First, on the familiarity with emotional AI, when the students are asked to choose the most appropriate definition of this technology, to the best of their knowledge, nearly 80% chose intelligent machine/algorithms that attempt to read (44.7%) or display (34%) the emotional state of humans. This means nearly 78% chose the roughly correct definitions of emotional AI and affective computing (McStay, 2018; Richardson, 2020; Rukavina et al., 2016). Meanwhile, 21.3% of the respondents chose AI that displays human consciousness (Figure 1A).

What is the best definition of emotional AI according to your knowledge?
1,015 responses

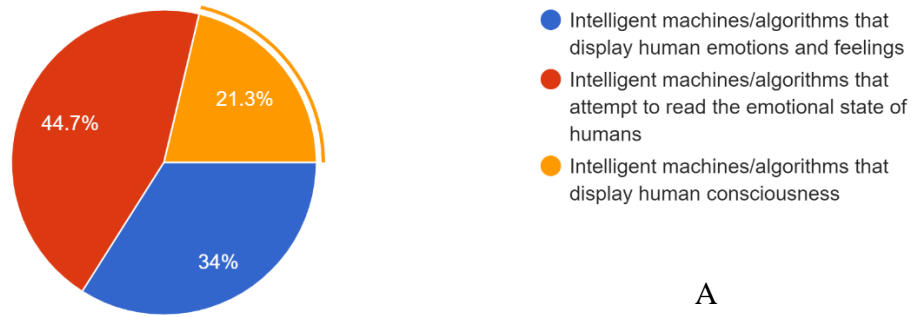

- Intelligent machines/algorithms that display human emotions and feelings
- Intelligent machines/algorithms that attempt to read the emotional state of humans
- Intelligent machines/algorithms that display human consciousness

A

How familiar are you with the topic of emotional AI
1,015 responses

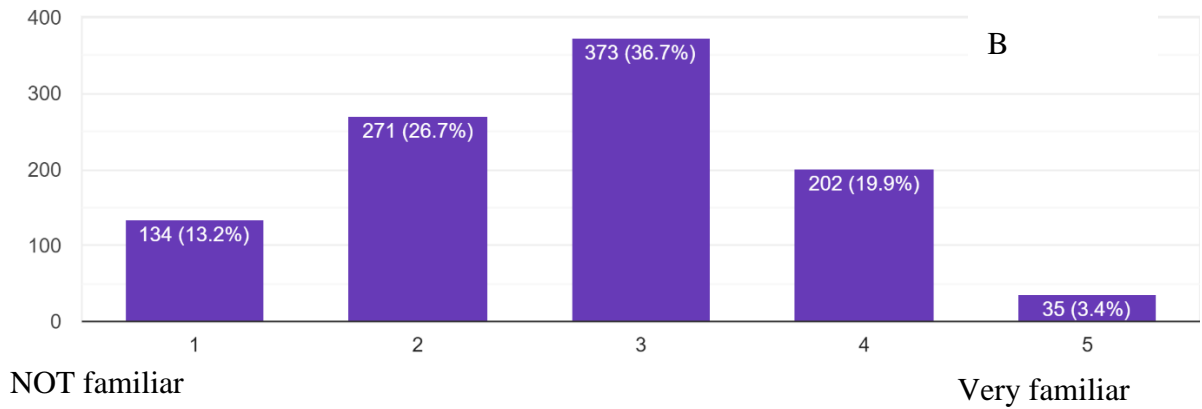

B

NOT familiar                                    Very familiar

**Figure 1:** Familiarity of the respondents with emotional AI. A) Students choose among three definitions of emotional AI. B) Students rate their familiarity with the topic.

In Figure 1B, when students are asked to rate their level of familiarity with emotional AI, there are around 23% who rate themselves as familiar to very familiar. Meanwhile, around 40% rate themselves as unfamiliar. These numbers show that although 80% of the respondents (Figure 1A) chose a very close definition of emotional AI; yet, the topic remains ambiguous for nearly the same percentage (Figure 1B).

In terms of ranking the ethical issues concerning AI, we presented students with a list of nine ethical problems with AI proposed by the World Economic Forum (Bossman, 2016) and asked them to choose topic three. Interestingly, the top concern for international students' body is essentially about human-machine interaction, i.e., "Humanity. How do machines affect our behavior and interaction?" with 561 responses (55.3%). The second greatest concern, at 488 responses or 48.1%, is about the security of these smart systems, i.e., "how do we keep AI safe from adversaries?". The third-place, surprisingly, is about unemployment with 467 responses or 46%. In this regard, this study converges with the academic literature on the perception of AI integration at work in that people are not very concerned about being replaced by AI but more concerned about human-AI interaction (Pinto dos Santos et al., 2019; Sarwar et al., 2019; Shen et al., 2020; Sit et al., 2020).

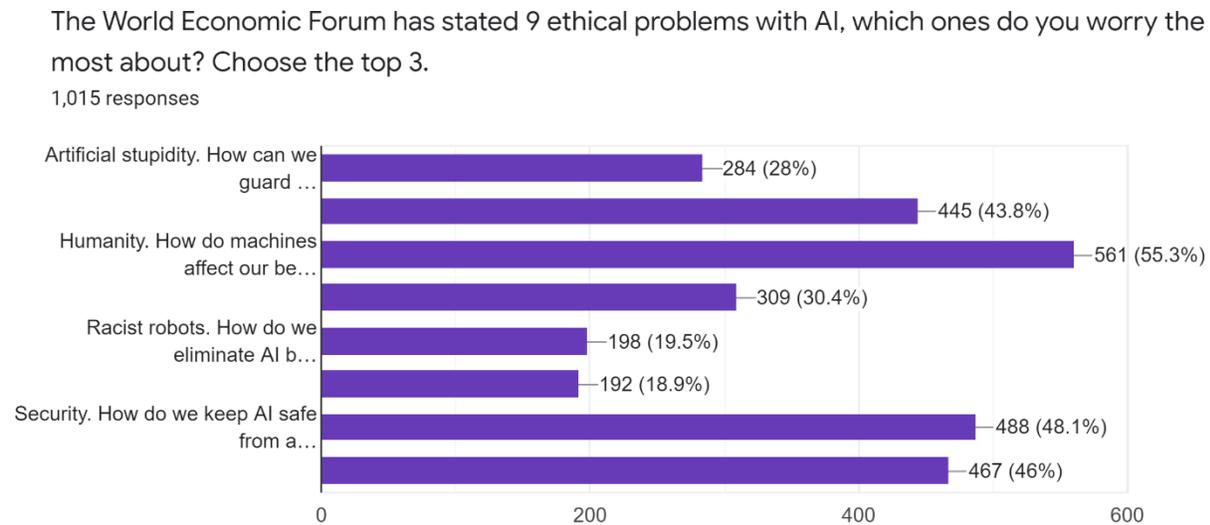

**Figure 2:** Nine ethical concerns regarding AI ranked by the students.

**Table 3:** Key characteristics of the surveyed sample

| Variables | Category | Male (N = 437) | | Female (N = 578) | |
|---|---|---|---|---|---|
| | | Frequency | Percentage | Frequency | Percentage |
| **Region** | Africa | 5 | 1.14% | 6 | 1.04% |
| | Central Asia | 11 | 2.52% | 5 | 0.87% |
| | Eastern Asia | 224 | 51.26% | 262 | 45.33% |
| | Europe | 9 | 2.06% | 11 | 1.90% |
| | Northern America | 7 | 1.60% | 10 | 1.73% |
| | South-Eastern Asia | 137 | 31.35% | 226 | 39.10% |
| | Southern Asia | 41 | 9.38% | 48 | 8.30% |
| | Oceania | 2 | 0.46% | 8 | 1.38% |
| **Income** | Low | 39 | 8.92% | 43 | 7.44% |
| | Medium | 327 | 74.83% | 483 | 83.56% |
| | High | 71 | 16.25% | 52 | 9.00% |
| **School year** | First year | 63 | 14.42% | 66 | 11.42% |
| | Second year | 118 | 27.00% | 198 | 34.26% |
| | Third year | 128 | 29.29% | 186 | 32.18% |
| | Fourth year | 111 | 25.40% | 109 | 18.86% |
| | Fifth year or more | 11 | 2.52% | 9 | 1.56% |
| **Major** | Business Management and Economics | 233 | 53.32% | 185 | 32.01% |
| | Social Sciences and Humanities | 204 | 46.68% | 392 | 67.82% |
| **Religions** | Atheism | 132 | 30.21% | 157 | 27.16% |
| | Buddhism | 64 | 14.65% | 129 | 22.32% |
| | Christianity | 59 | 13.50% | 66 | 11.42% |
| | Islam | 52 | 11.90% | 58 | 10.03% |
| | Others or Unidentified | 130 | 29.75% | 168 | 29.07% |
| **Religiosity** | Mildly religious & Not religious | 372 | 85.13% | 494 | 85.47% |
| | Very religious | 36 | 8.24% | 45 | 7.79% |
| **Attitude toward the AI managers** | average to positive | 243 | 55.61% | 244 | 42.21% |
| | negative to average | 194 | 44.39% | 334 | 57.79% |
| **Familiarity with AI** | Above average | 258 | 59.04% | 238 | 41.18% |
| | Below average | 179 | 40.96% | 340 | 58.82% |

Table 3 shows 52% of the respondents hold a negative view of the AI managers, and 51% rated themselves below average regarding AI knowledge.

***Technical validation***

*Convergence diagnostics*

After running the MCMC analyses for all 10 models (4 chains, 5000 iterations, 2000 warm-ups), the basic two standard diagnostic tests return good results. All ***Rhat***'s values equal one (1), and all the effective sample sizes (***n_eff***) are above 1000. The detailed results and visualizations (of the autocorrelation coefficient, the Gelman Shrink Factor, the Markov chains) of the diagnostic tests are presented in the Supplementary folder.

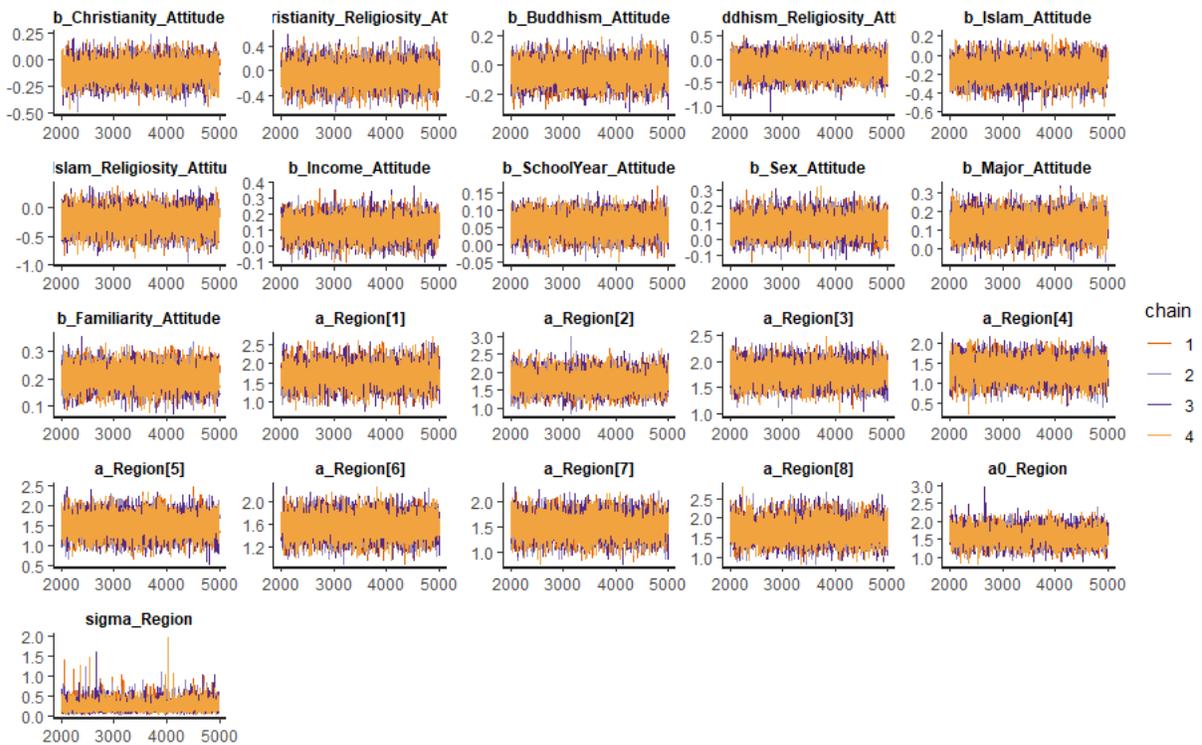

**Figure 3:** The mixing of the Markov chains after fitting the model.

*Model comparison*

First, after fitting the models, we run the PSIS-LOO test for all of the models and find that all Pareto k estimates are good (k<0.5) (See more details in the Supplementary Folder). In Bayesian

|  | Pseudo-BMA without Bayesian bootstrap | Pseudo-BMA with Bayesian bootstrap | Bayesian stacking |
|---|---|---|---|
| Model 1 | 0.000 | 0.005 | 0.056 |
| Model 2 | 0.000 | 0.001 | 0.112 |
| Model 3 | 0.000 | 0.000 | 0.000 |
| Model 4 | 0.000 | 0.001 | 0.000 |
| Model 9 | 0.001 | 0.068 | 0.000 |

statistics, the weight is calculated to distribute the level of plausibilities among different models given the data. We compare all the models using for types of weights as shown in Table 4

. **Table 4:** Weight comparison among model with *Attitude* as the outcome variable

| | | | |
|---|---|---|---|
| Model 10 | 0.999 | 0.924 | 0.833 |

Table 4 shows that model 10 starkly outperforms all other models in all categories of weight, out of all the models with attitude toward the AI managers as the outcome variable. Meanwhile, table 5 shows that model 5 fits the data the best among models with self-rated familiarity with emotional AI as the outcome variable.

**Table 5:** Weight comparison among model with *Familiarity* as the outcome variable

| | Pseudo-BMA without Bayesian bootstrap | Pseudo-BMA with Bayesian bootstrap | Bayesian stacking |
|---|---|---|---|
| Model 5 | 0.821 | 0.672 | 0.685 |
| Model 6 | 0.000 | 0.030 | 0.083 |
| Model 7 | 0.000 | 0.016 | 0.057 |
| Model 8 | 0.179 | 0.282 | 0.175 |

The visualization of model 5 and 10 are presented in Figure 4 below:

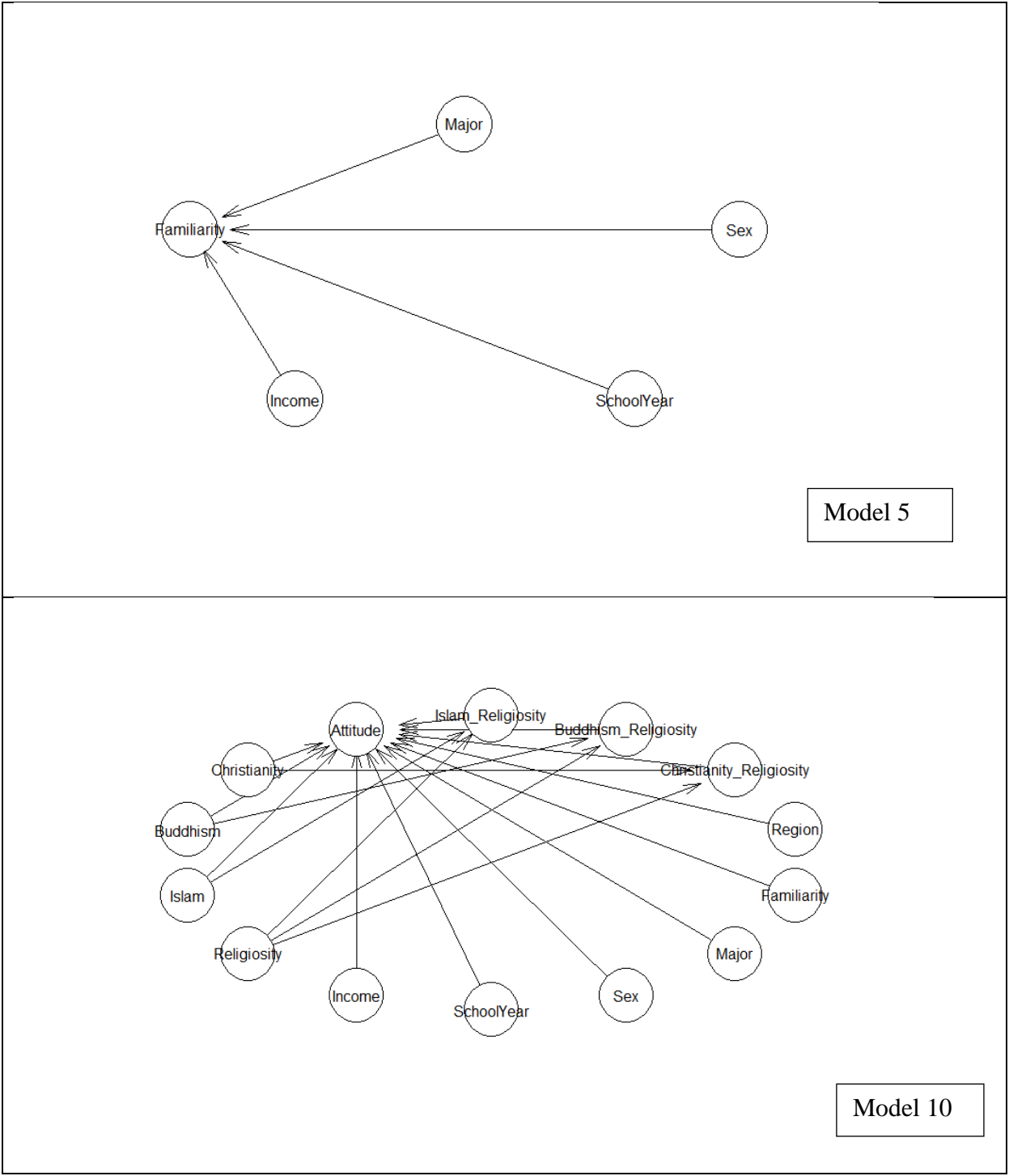

**Figure 4:** Visualizations of model 5 and 10, the two models with the highest goodness-of-fit.

**Table 6:** A robustness check on the prior sensitivity for model 10

| **Priors:** b_Familiarity_Attitude ~ normal (-1,10) | | | | |
|---|---|---|---|---|
| | **Mean** | **SD** | **n_eff** | **Rhat** |
| a_Attitude | 1.61 | 0.21 | 5742 | 1 |
| b_Christianity_Attitude | -0.10 | 0.09 | 10326 | 1 |
| b_Christianity_Religiosity_Attitude | -0.05 | 0.17 | 12154 | 1 |
| b_Buddhism_Attitude | -0.05 | 0.07 | 11641 | 1 |
| b_Buddhism_Religiosity_Attitude | -0.15 | 0.19 | 12145 | 1 |
| b_Islam_Attitude | -0.16 | 0.11 | 10245 | 1 |
| b_Islam_Religiosity_Attitude | -0.24 | 0.17 | 10254 | 1 |
| b_Income_Attitude | 0.12 | 0.06 | 10254 | 1 |
| b_SchoolYear_Attitude | 0.06 | 0.03 | 11245 | 1 |
| b_Sex_Attitude | 0.09 | 0.06 | 13245 | 1 |
| b_Major_Attitude | 0.13 | 0.06 | 11201 | 1 |
| b_Familiarity_Attitude | 0.21 | 0.04 | 10154 | 1 |
| a_Region[Africa] | 1.70 | 0.26 | 9343 | 1 |
| a_Region[Central Asia] | 1.75 | 0.25 | 9371 | 1 |
| a_Region[Eastern Asia] | 1.78 | 0.19 | 7333 | 1 |
| a_Region[Europe] | 1.36 | 0.25 | 9355 | 1 |
| a_Region[Northern America] | 1.53 | 0.25 | 8539 | 1 |
| a_Region[South-Eastern Asia] | 1.57 | 0.19 | 8091 | 1 |
| a_Region[Southern Asia] | 1.51 | 0.20 | 8421 | 1 |
| a_Region[Oceania] | 1.70 | 0.25 | 8245 | 1 |
| *Monte Carlo SE of elpd_loo is 0.0.* | | | | |
| *All Pareto k estimates are good (k < 0.5).* | | | | |
| **Priors:** b_Familiarity_Attitude ~ normal (1,10) | | | | |
| | **Mean** | **SD** | **n_eff** | **Rhat** |

| | | | |
|---|---|---|---|
| a_Attitude | 1.61 | 0.21 | 5221 | 1 |
| b_Christianity_Attitude | -0.10 | 0.09 | 9421 | 1 |
| b_Christianity_Religiosity_Attitude | -0.05 | 0.17 | 10144 | 1 |
| b_Buddhism_Attitude | -0.05 | 0.07 | 10879 | 1 |
| b_Buddhism_Religiosity_Attitude | -0.15 | 0.19 | 12547 | 1 |
| b_Islam_Attitude | -0.16 | 0.11 | 10254 | 1 |
| b_Islam_Religiosity_Attitude | -0.24 | 0.17 | 11245 | 1 |
| b_Income_Attitude | 0.12 | 0.06 | 10544 | 1 |
| b_SchoolYear_Attitude | 0.05 | 0.03 | 12584 | 1 |
| b_Sex_Attitude | 0.09 | 0.06 | 125448 | 1 |
| b_Major_Attitude | 0.13 | 0.06 | 13121 | 1 |
| b_Familiarity_Attitude | 0.21 | 0.04 | 12421 | 1 |
| a_Region[Africa] | 1.70 | 0.26 | 9511 | 1 |
| a_Region[Central Asia] | 1.75 | 0.24 | 8816 | 1 |
| a_Region[Eastern Asia] | 1.78 | 0.18 | 9421 | 1 |
| a_Region[Europe] | 1.35 | 0.25 | 8147 | 1 |
| a_Region[Northern America] | 1.53 | 0.25 | 7620 | 1 |
| a_Region[South-Eastern Asia] | 1.58 | 0.19 | 9245 | 1 |
| a_Region[Southern Asia] | 1.51 | 0.20 | 8651 | 1 |
| a_Region[Oceania] | 1.70 | 0.26 | 9453 | 1 |

*Monte Carlo SE of elpd_loo is 0.0.*

*All Pareto k estimates are good (k < 0.5).*

Table 6 shows the tweaking of the Bayesian priors results in no real differences in the posterior distribution, suggesting the model is robust.

### *Major findings*

#### *The multifacetedness of attitude toward the AI managers*

The best performances were shown by model 10 and model 5. The results demonstrate that attitude toward the AI managers is a very multi-faceted issue. It is determined by socio-demographic factors and culturally and politically related factors such as religions and religiosity, and region. Self-rated familiarity with the topic of emotional AI, however, is less complicated. It can be predicted from basic socio-demographic factors such as sex, school year, income, and major, as shown in model 5. In the next section, we will compare the posterior distribution from multiple models to find consistent results across all models. However, as Model 10 and Model 5 fit the data the best, their findings will tend to be more reliable.

#### *Determinants of self-familiarity with AI*

**Table 7:** Posterior distribution of variables in model 5

|  | **Mean** | **SD** | **n_eff** | **Rhat** |
|:---:|:---:|:---:|:---:|:---:|
| a_Familiarity | 1.93 | 0.17 | 7852 | 1 |
| b_Income_Familiarity | 0.03 | 0.06 | 7156 | 1 |
| b_SchoolYear_Familiarity | 0.01 | 0.03 | 10521 | 1 |
| b_Sex_Familiarity | 0.25 | 0.06 | 10925 | 1 |
| b_Major_Familiarity | 0.10 | 0.06 | 12852 | 1 |
| *Monte Carlo SE of elpd_loo is 0.0.* <br> *All Pareto k estimates are good (k < 0.5).* | | | | |

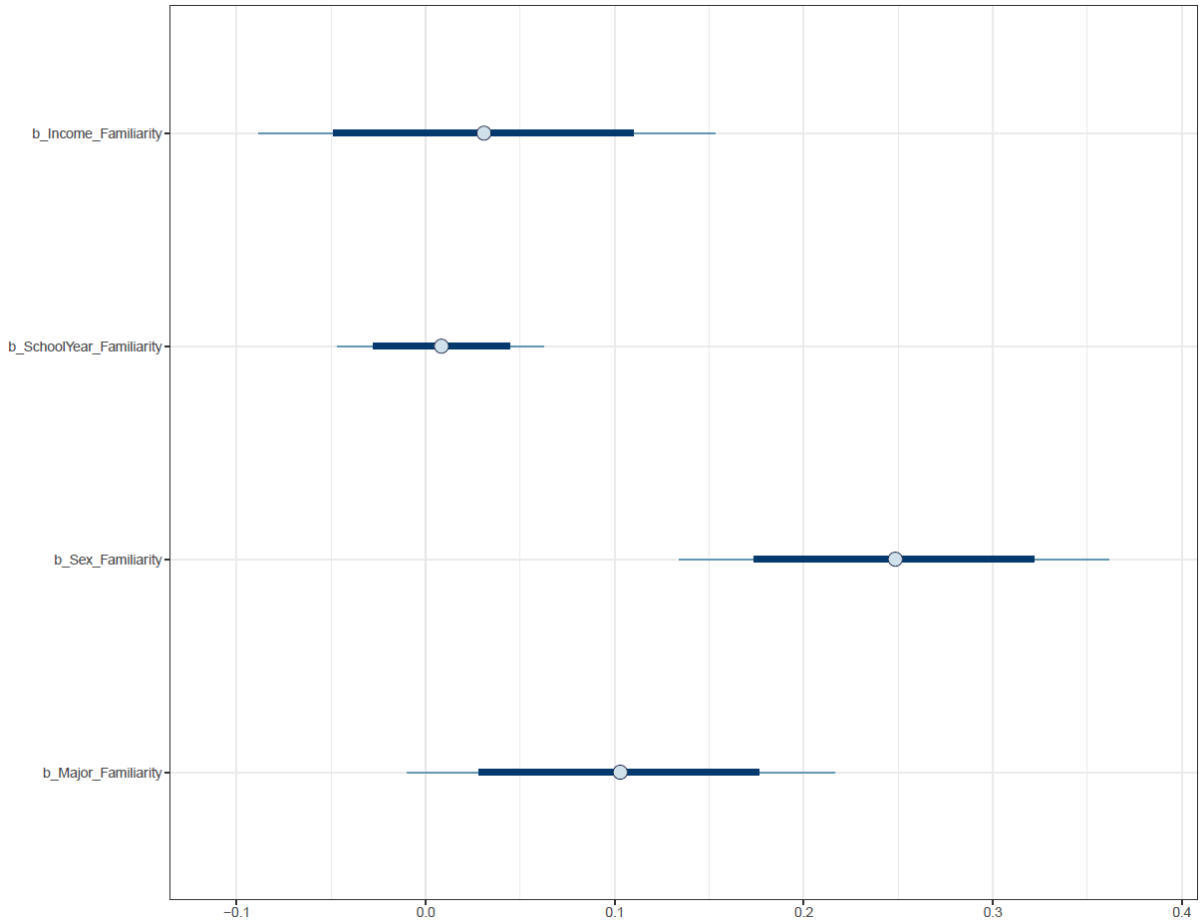

**Figure 5:** Interval plot of the posterior distribution of income, school year, sex, and major to predict self-familiarity with AI.

As shown in Figure 5 and Table 7, major and sex are the most predictive of self-rated familiarity with AI, while income and school year are not. Specifically, being male and having a business management major are predictive of rating oneself as more familiar with AI. Since the male gender has been found to positively associate with perceived technological self-efficacy, i.e., the belief that one is capable of performing a task using technologies (Cai et al., 2017; Huffman et al., 2013; Mackay & Parkinson, 2010; Tømte & Hatlevik, 2011; Vekiri & Chronaki, 2008), this study's finding on the correlation of male gender and higher reporting of familiarity with AI confirms this well-established trend. While the effects of the school year and income are more ambiguous, given the posterior distribution.

*Determinants of the general attitude toward the AI managers*

Table 8 below presents the posterior distribution of all variables of model 10, which show the most goodness-of-fit among models with attitude as the outcome variable.

**Table 8:** Posterior distribution of variables in model 10.

|  | Mean | SD | n_eff | Rhat |
|---|---|---|---|---|
| a_Attitude | 1.61 | 0.35 | 6744 | 1 |
| b_Christianity_Attitude | -0.10 | 0.09 | 9854 | 1 |
| b_Christianity_Religiosity_Attitude | -0.05 | 0.17 | 10445 | 1 |
| b_Buddhism_Attitude | -0.05 | 0.07 | 10545 | 1 |
| b_Buddhism_Religiosity_Attitude | -0.15 | 0.19 | 10186 | 1 |
| b_Islam_Attitude | -0.16 | 0.10 | 9251 | 1 |
| b_Islam_Religiosity_Attitude | -0.24 | 0.18 | 9074 | 1 |
| b_Income_Attitude | 0.12 | 0.06 | 9551 | 1 |
| b_SchoolYear_Attitude | 0.06 | 0.03 | 11553 | 1 |
| b_Sex_Attitude | 0.09 | 0.06 | 9601 | 1 |
| b_Major_Attitude | 0.13 | 0.06 | 12210 | 1 |
| b_Familiarity_Attitude | 0.21 | 0.04 | 9532 | 1 |
| a_Region[Africa] | 1.70 | 0.26 | 6458 | 1 |
| a_Region[Central Asia] | 1.75 | 0.24 | 10214 | 1 |
| a_Region[Eastern Asia] | 1.78 | 0.18 | 9138 | 1 |
| a_Region[Europe] | 1.36 | 0.26 | 8058 | 1 |
| a_Region[Northern America] | 1.53 | 0.24 | 8558 | 1 |
| a_Region[South-Eastern Asia] | 1.58 | 0.19 | 9271 | 1 |
| a_Region[Southern Asia] | 1.51 | 0.20 | 8872 | 1 |
| a_Region[Oceania] | 1.70 | 0.26 | | |
| *Monte Carlo SE of elpd_loo is 0.0.* | | | | |
| *All Pareto k estimates are good (k < 0.5).* | | | | |

Sex, income, school year, major, familiarity

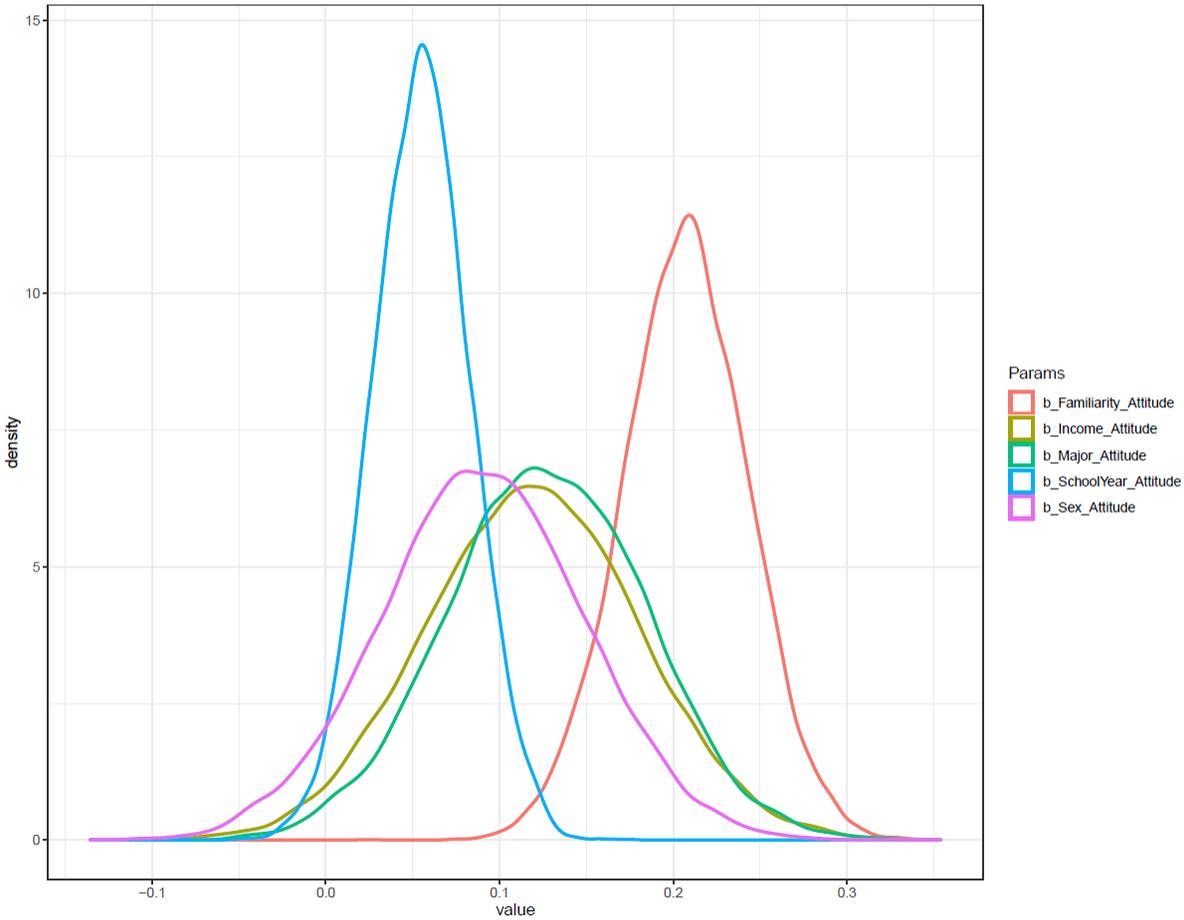

**Figure 6:** Density plot from Model 10 for five variables: familiarity, income, major, school year, and sex

Across Model 1, Model 4, Model 9, and Model 10, the results consistently show students with higher income, being a male, business major, and being a senior are likely to have a less-worrying outlook toward AI-managers. Regarding income, one of the possible explanations might be the students with higher income are likely to have higher educational attainment (Aakvik et al., 2005; Blanden & Gregg, 2004) and end up in high-status occupations (Macmillan et al., 2015); thus they are less worried about being managed by AI algorithms. In all likelihood, they are more likely to become future managers who will use those AI tools to recruit and monitor their employees.

Regarding the sex variables, our result is aligned with the well-established result in the literature that male-ness is correlated with higher perceived technological self-efficacy (Huffman et al., 2013; Mackay & Parkinson, 2010; Tømte & Hatlevik, 2011; Vekiri & Chronaki, 2008).

Being a business major is correlated with less anxiety toward AI applications in human resource management might be a product of the lack of emphasis on AI's ethical and social implications in business education or the business students' focus on making their future businesses profitable. It is reasonable to assume that these business majors' hope is to one day end up in a managerial position. Being a manager would incline a person to adopt the company position, thus seeing things in terms of productivity and performance results and not questioning (Ball, 2010).

Being a senior is correlated with a less worrying outlook of the AI managers could be interpreted that the concern of the students who are near their graduation is more on finding a job and less about any potential bias or misuse in the future application of AI in the workplace.

Model 10 shows that students who rate themselves as more familiar with emotional AI tend to view the application of this technology in the workplace more positively. This result contradicts a Saudi Arabian study of medical students (Bin Dahmash et al., 2020), which found anxiety toward using AI was correlated with a higher self-perceived understanding of this technology. This divergence with the literature can be explained by the diversity of the surveyed population, of which there are 48 countries and 8 regions. Next, we will investigate the influence of religions and religiosity on the perception of AI human resource managers.

Religions and religiosity

Buddhist students are least likely to have a more worrying outlook toward the non-human bosses. While Muslim students are most likely to have a negative attitude, the coefficient ***b_Islam_Attitude*** (mean = -0.10, sd = 0.09) is distributed mostly on the negative side. Christian students are more ambiguous, but the majority of the ***b_Christianity_attitude***'s distribution is on the negative side (mean = -0.10, sd = 0.09). These trends are visualized in Figure 7.

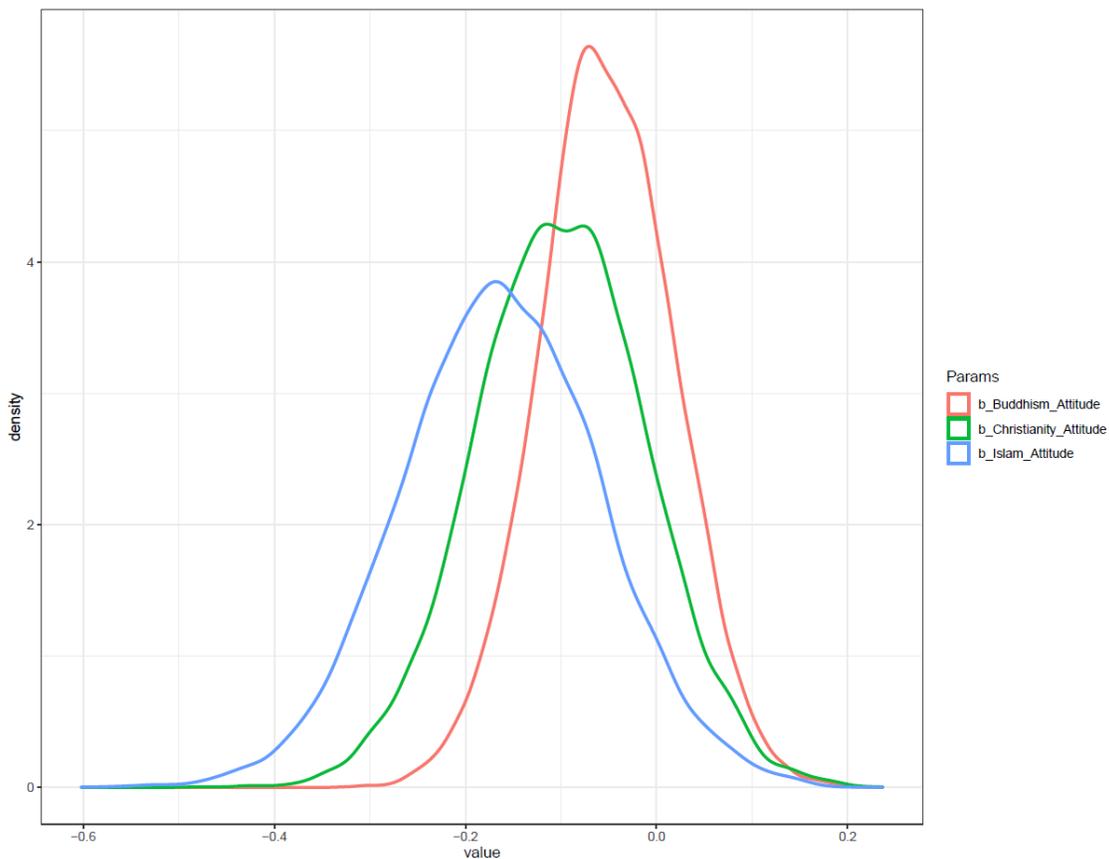

**Figure 7:** The density plot of the *Religion* variable

Higher religiosity of the Muslim and Buddhist students appears to have made the students more anxious about using AI in human resource management. In Table 8, ***β_Islam_Attitude*** (mean =-0.16, sd =0.10); ***β_Islam_Religiosity_Attitude*** (***mean =*** -0.24, sd = 0.18). There is a similar trend for the Buddhist students as well with ***β_Buddhism_Attitude***'s mean =-0.05, sd =0.07; ***β_Buddhism_Religiosity_Attitude's*** mean **=** -0.15, sd = 0.19).

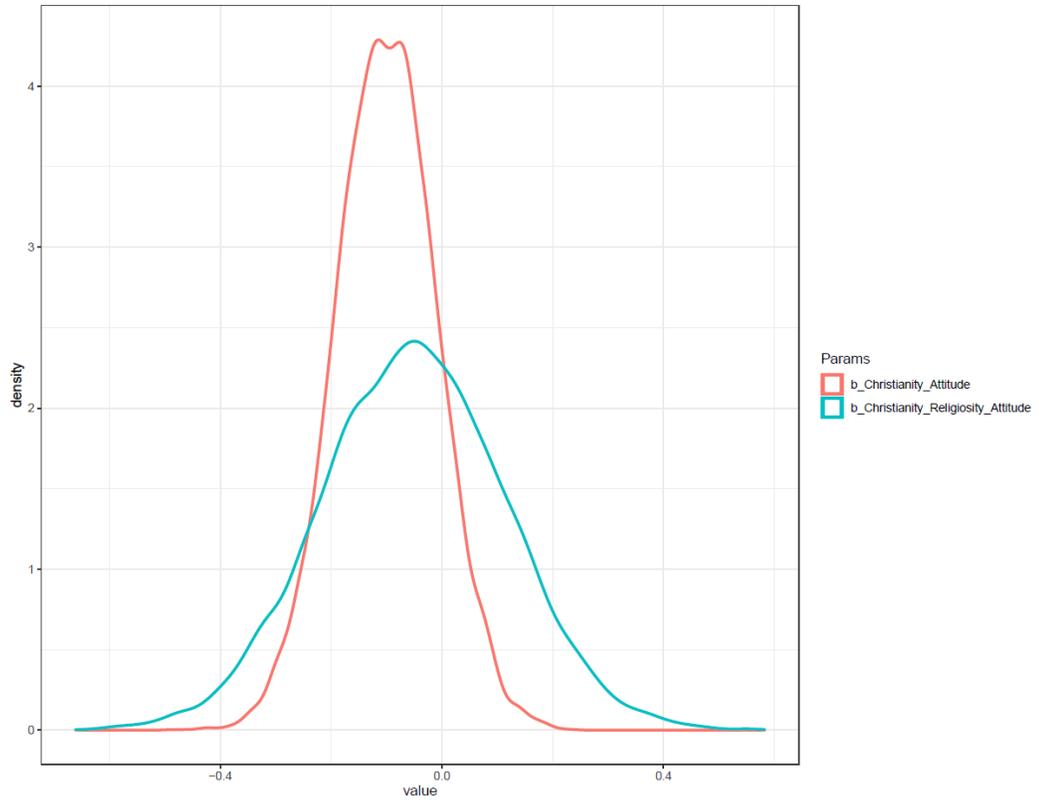

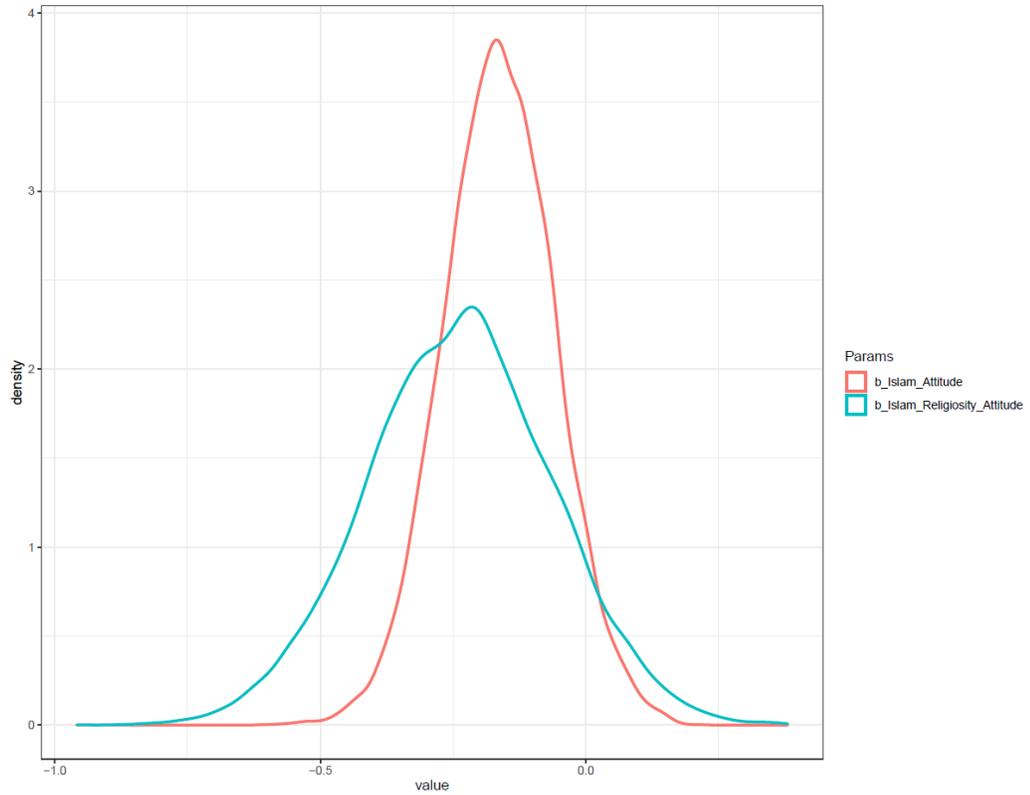

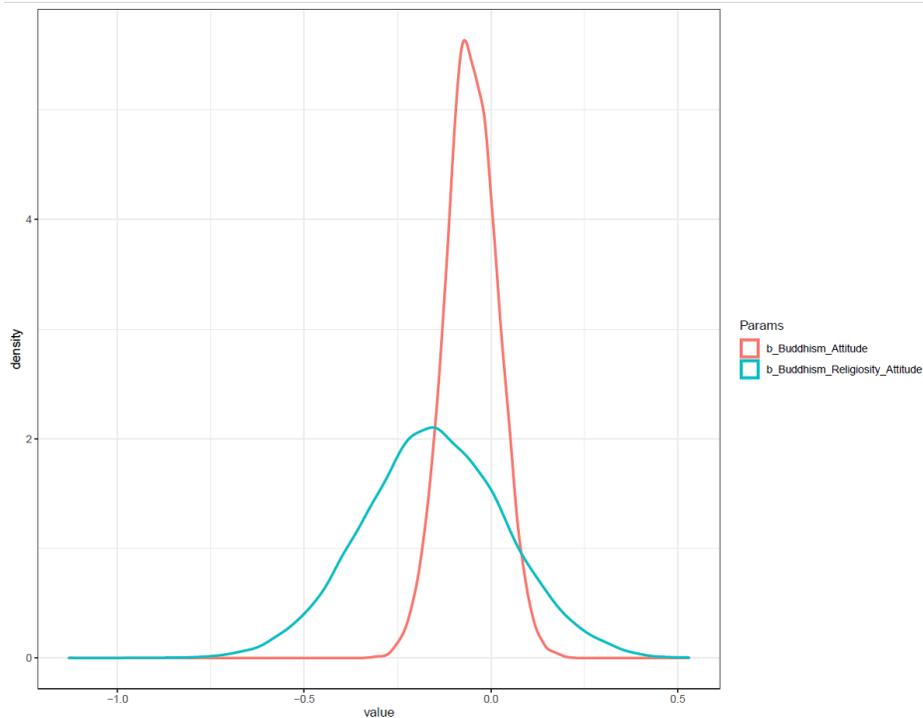

**Figure 8:** Density plot of the Buddhism, Islam, and Christianity variable

However, the Christian respondents' high religiosity seems to generate a slight shift of the distribution toward the positive range and makes the distribution wider. The mean value of **β_Christian_Attitude** is -0.10 (sd= 0.09), while the mean value of **β_Christianity_Religiosity_Attitude** is -0.05 (sd= 0.17). These trends are visualized in Figure 7 and 8.

The finding of being a Muslim and being a highly religious Muslim is correlated with a more worrying outlook with the AI managers is expected as Muslim and Islam are portrayed negatively in the media, as shown in a meta-analysis of relevant studies from 2005 to 2016 by Ahmed and Matthes (2016). They are also subjected to various forms of discrimination, real and perceived, in a post 9/11 world. On the other hand, Buddhism or Christianity as one's religion is also correlated with higher anxiety toward non-human resource management suggests having an official religion at all might make the respondents more sensitive to the potential misuse and bias in AI applications in the modern human resource management context.

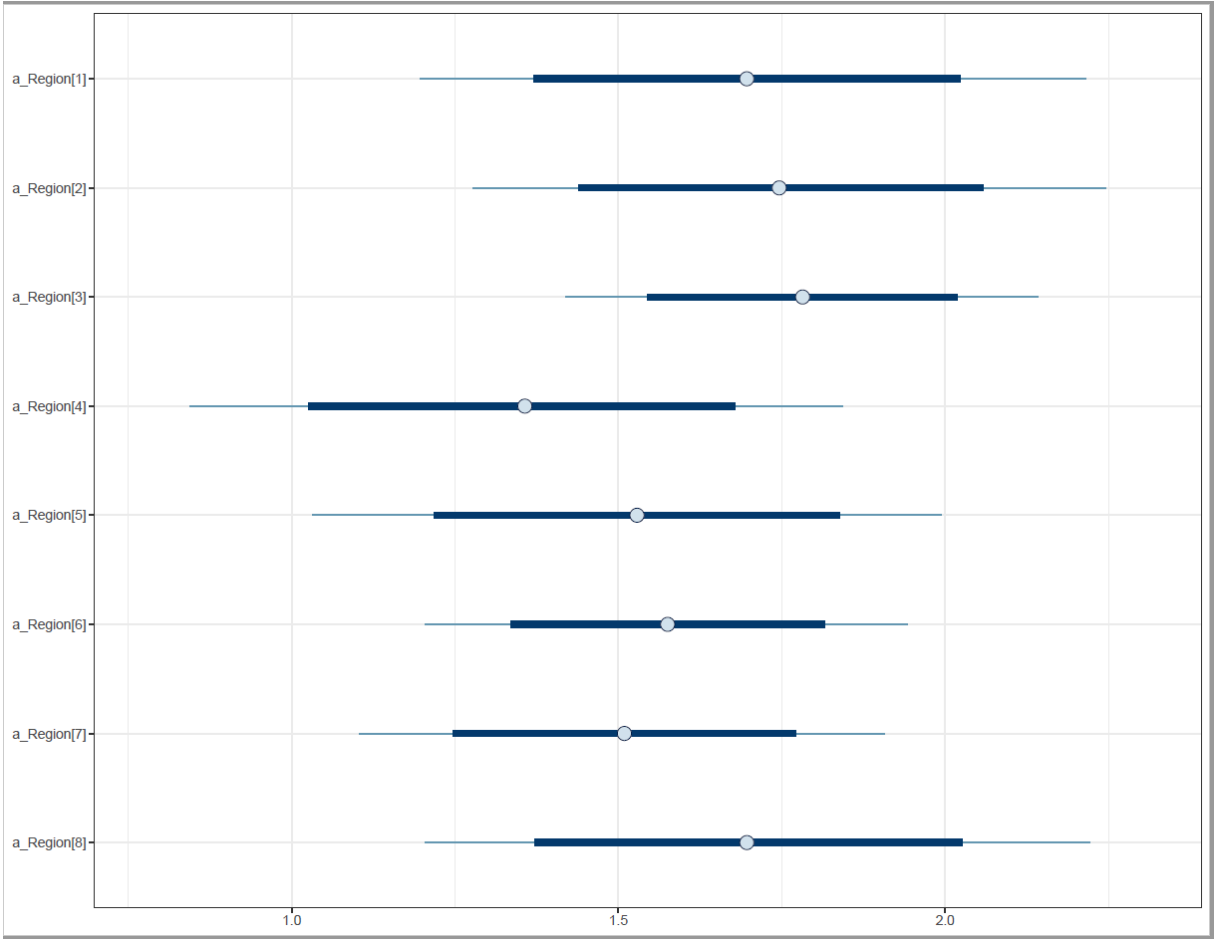

**Figure 9:** Interval plot of the Region variable: 1) Africa; 2) Central Asia; 3) Eastern Asia; 4) Europe; 5) Northern America; 6) South-Eastern Asia; 7) Southern Asia; 8) Oceania

The attitudes toward AI human-resource management of the respondents from the different geographical regions are also different. Respondents from Eastern Asia have the lowest anxiety (***a_Region[Eastern Asia]*** =1.78; sd = 0.18), while respondents from Europe are the most likely to be wary of the use of non-human resource management (***a_Region[Europe]*** = 1.36; sd

= 0.26). Such finding might be rooted in the cultural differences among the regions. For example, there are well-established results from the psychology literature about the differences between collectivist cultures and individualist cultures (de Oliveira & Nisbett, 2017; Henrich, 2020). In the collectivist culture, for example, in Eastern Asia typically (Japan, China, South Korea, etc.), concerns about privacy and self-autonomy are less pronounced compared with the individualist counterpart (Henrich, 2020; Whitman, 1985). Legitimate studies on the perception of AI tools in "moral governance" in China, for example, are rather positive (Roberts et al., 2020). Such cultural differences could explain why Eastern Asian respondents express the least anxiety toward the prospect of AI managing the job-screening and work-monitoring tasks.

Another notable result is that students from underdeveloped regions (Africa, Central Asia, Oceania) also tend toward a lower level of anxiety toward being managed by AI (Table 8 and Figure 9). This is likely the result of a concern with job finding and economic development over privacy and self-autonomy, predominant in AI discourses in developed, Western countries (Zuboff, 2019).

**Discussion and Conclusion**

***Limitations and future research directions***

This study suffers from several limitations. First, it inherits the limitations of the convenient sampling method. First, as the surveyed population is young students who study in a multicultural, bilingual campus (Nguyen et al., 2019), the results should be interpreted in that context. Second, some regions such as Eastern Asia and South-Eastern Asia are over-represented in the sample. However, with a large and diverse sample of 1,015 respondents from 48 countries and 8 regions, the paper still makes a good contribution to the AI-perception literature, which is currently skewed toward the country and profession-specific findings.

As this is among a few cross-cultural studies that focus on applying AI in human resource management, future studies can further explore the causal mechanisms of the correlations established in this study. For example, conducting in-depth interviews and controlled experiments with respondents from diverse cultural backgrounds can explain the influences of religions and religiosity on AI's perception.

***Implications***

This study has several contributions to the literature. Besides the fact that it is among the few cross-cultural empirical studies on the perception of the use of AI in human resource management, the paper discovers that being managed by AI is the greatest AI risk perceived by the international future job seekers. Moreover, the analytical insights highlight the urgent need for better education and science communication about AI's risk in the workplace. The cross-cultural and socio-demographic discrepancies in concern and ignorance about the AI managers can be bridged. Finally, methodologically, the use of Bayesian multi-level modeling shows a great advantage of the traditional frequentist multivariate regression in quantifying the regional and religious correlates of AI perception.

*Being managed by AI is the greatest cause for concern*

The descriptive statistics section has indicated that being managed by AI and interaction with AI are major concerns for the respondents. Table 3 shows that 52% of the future job seekers express negative concern about the AI managers on average. In contrast, Figure 2 shows that human-AI interaction, i.e., "How do machines affect our behavior and interaction?" is the most outstanding ethical concern for the students with nearly 55% of the total responses. In comparison, job loss to AI only ranks third with 48%. These insights will prove crucial when communicating about the risks of AI. As workplace surveillance seeks to go beyond the exterior of the physical body and attempts to datafy their emotional lives (Ball, 2010; Marciano, 2019; Richardson, 2020), the greatest worry for young jobseekers is not AI replacing their jobs rather AI supervising, evaluating and making decisions about their performance and career advancement. This is perhaps the most significant concern from a majority of our respondents. In the next section, the implications and nuances of the cross-cultural and socio-demographic correlate for the perception of AI use in human resource management are discussed.

*Bridging the cross-cultural discrepancies*

As highlighted in the section on regional and religious differences, people from different socio-cultural, economic backgrounds tend to form different perceptions of emerging technologies. Here, it is worth recalling previous studies on workplace surveillance, which show the employees' awareness of the presence of the smart surveillance technologies negatively correlates with organizational commitment (Ball, 2010; Brougham & Haar, 2017). These two tendencies, combined with the risk of AI being misunderstood (Wilkens, 2020), are important obstacles to overcome to harness the technologies' potential for good.

In terms of regional differences, our analysis shows people from economically less developed regions (Africa, Oceania, Central Asia) exhibit less concern for AI managers (See Figure 9), while those surveyed from more economically prosperous regions (Europe, Northern America) tend to be more cautious and reserved. However, it is interesting to note that an economically prosperous region such as East Asia (including China, Japan, Mongolia, North Korea, South Korea, and Taiwan) correlates with less anxiety toward the AI managers.

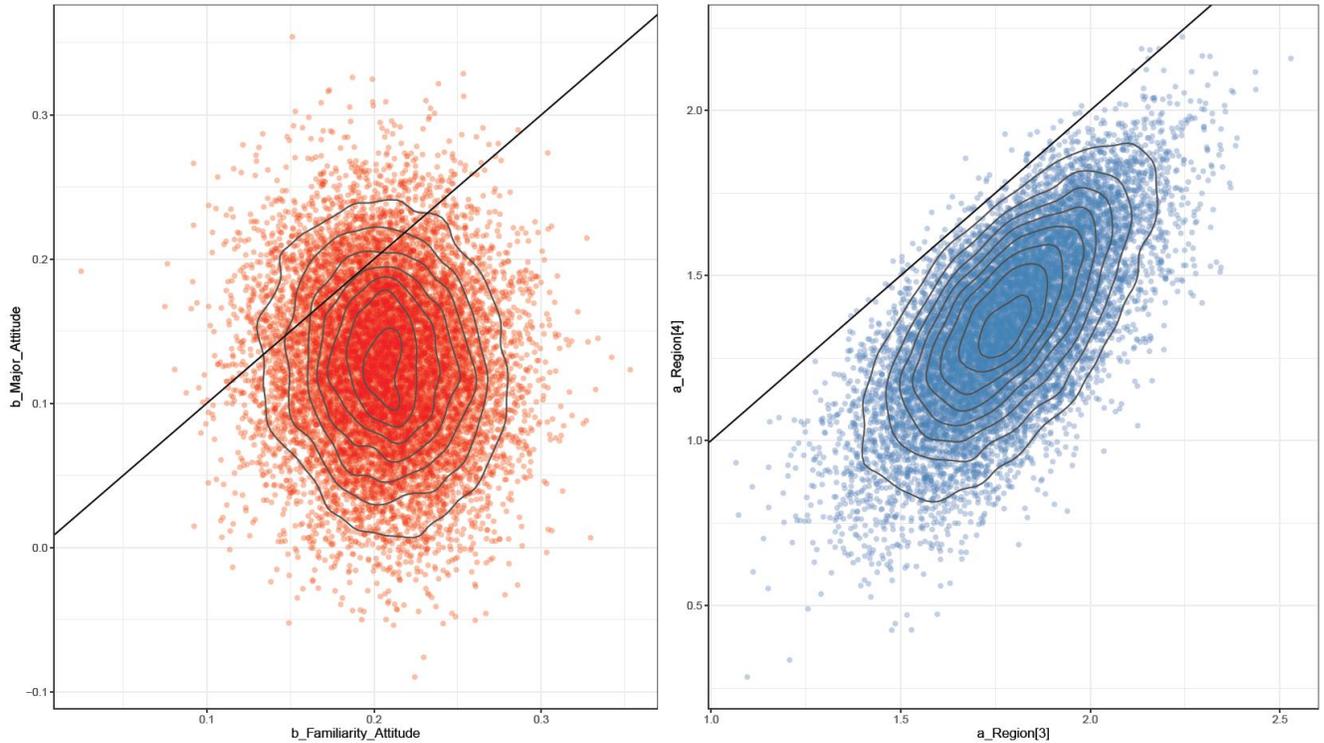

**Figure 10:** A chart of simulated values of a striking pair of coefficients: ***b_Major_Attitude*** vs. ***b_Familiarity_Attitude*** on the left, and ***a_Region[3]*** (Eastern Asia) vs. ***a_Region [4]*** (Europe).

Our data show the opposite trends in the perception toward the AI managers between European/North American and people from three major East Asian countries. In the latter case, more people are showing an accepting attitude rather than a worrying or neutral attitude: 64% of the Japanese, 56% of the South Korean, and 42% of the Chinese exhibit the accepting attitude. For European and Northern Americans, an overwhelming majority of 75% possess the worrying/neutral attitude toward being managed by AI (see Figure 10).

**Table 9**: Comparing the distribution of different attitudes toward the AI managers by three major East Asian countries (China, Japan, Korea) and Europe/North America

| Countries/Regions | Attitude toward AI managers | Frequency | Percentage |
|---|---|---|---|
| Chinese | Worrying/Neutral | 46 | 58% |
| | Accepting | 33 | 42% |
| Japanese | Worrying/Neutral | 103 | 36% |
| | Accepting | 180 | 64% |
| South Korea | Worrying/Neutral | 38 | 44% |
| | Accepting | 49 | 56% |
| European/North American | Worrying/Neutral | 24 | 75% |
| | Worrying/Neutral | 8 | 25% |

Since these East Asian countries have different political systems, the highly accepting attitude could be explained by Confucianism's influence on the way people perceive moral issues such as privacy, work ethics, harmony, duty, and loyalty. Specifically, there might be an antipathy toward rights in Confucian culture (Chung et al., 2008; Weatherley, 2002), and there is a stronger emphasis on harmony, duty, and loyalty to the collective will (Vuong et al., 2018; Vuong et al., 2020). Moreover, in Confucian ethics, the ultimate goal of self-development is to transcend one's self-interests and to align with things that are bigger than oneself: one's own community, nation, and nature (Nguyen & Ho, 2019; Whitman, 1985). Moreover, there is much more acceptance of moral intervention as people in the higher position of the social hierarchy and the government, at their best, are usually thought of as a source of moral guidance (Roberts et al., 2020).

The empirical findings on such stark cross-cultural and cross-regional differences could help educators, businesses, and policymakers to shape their action programs to address any stakeholder's concern or lack thereof for the future of an AI-powered workplace (Condie & Dayton, 2020). For example, as shown in Table 8 and Figure 7, religious respondents in the sample tend to show a highly negative attitude toward using AI in human resource management. Such concern needs to be addressed. As shown in this study, 21% of the respondents still equate emotional AI with machines or algorithms that display human consciousness (Figure 1A). Nearly 40% rated themselves as unfamiliar with emotional AI (Figure 1B).

*Ignorance of biases and privileges*

Our analysis indicates that many students are ignorant of their own biases/privileges and possible harmful social and ethical implications of AI managers. For example, while this study shows that being male and being from a higher-income background are correlated with less anxiety regarding non-human resource management (Table 8 and Figure 6), it also reveals that being of a specific religious denomination, being female, raises serious concerns about algorithmically driven managers. As past studies have shown, student engagement with ethics is contingent on several factors: first, the type of curriculum adopted by higher education institutions (Culver et al., 2013); second, how the concept of bias is communicated and understood through the course literature. And finally, courses orientated toward technology, where the potential pitfalls of algorithmic bias found in many current AI applications (Azari et al., 2020; Buolamwini & Gebru, 2018) are taught problematically, are only made available to a privileged class who can attend post-secondary institutions.

On the latter, Table 8 and Figure 10 shows that self-rated familiarity with AI is correlated with a more positive outlook of the AI mangers (**β_Familiarity_Attitude's** mean = 0.21, sd = 0.04), which implies these students might be unaware of the biases and inaccuracy in the emerging technologies. Even though the problem of algorithmic bias has now moved from the periphery to the center of public discourse, as witnessed in Cathy O'Neil's New York Times bestseller, *Weapons of Math Destruction* (O'Neil, 2016), the proposed US Algorithmic Accountability Act, or the recent media storm surrounding the firing of Google AI's top ethics researcher, Timnit Gebru (Singh, 2020; Hao, 2020), when it comes to the cross-cultural, multi-national population, this study indicates a clear lack of knowledge (Table 3).

There seem to be little push-backs on the use of emotional AI technologies in East Asian countries to bring these insights into context. For example, in South Korea, nearly 25% of the top 131 corporations stated they were planning to facilitate their recruitment with emotional AI tools (Condie & Dayton, 2020). Large corporations such as Softbank, Honda, IBM, Microsoft, etc., are all pushing the production, promotion, and sales of these smart technologies. Under the influence of the COVID-19 pandemic, this trend is only growing. Many researchers have raised concerns regarding the use of AI technologies to surveil the population (Roussi, 2020) and the workforce (Condie & Dayton, 2020), which can linger well after the pandemic.

As such, university curriculums must include courses on social and ethical implications of AI in the workplace, especially in the business major, which has been shown to correlate with less concern about AI in HR management in this paper (see Figure 10). This is to correct any students' misconceptions and enrich their understanding of the positive (optimizing workplace productivity) and negative (algorithmic bias) potential of such technologies. It is important to remember that emotional AI technologies are designed to be *empathic*, not *sympathetic* managers. Evaluation systems such as STEM and university level courses in Information Technology or applied business practices are insufficient tools in preparing future job seekers for the quantified workplace. Rather, ethical training and critical thinking should be integral to institutional higher learning epistemology that prepares younger generations for an AI-augmented workforce.

**Conclusions**

Philosopher Toby Ord uses the metaphor of the precipice to help us visualize the existential risks humanity faces with the rise of smart technologies (Ord, 2020). The transformation of the modern workplace through algorithmic decisionism and governance prompts urgent questions that stakeholders need to address, from software engineers and programmers to policymakers and business leaders about how to live ethically and well with *'machines that feel.'* Yet, there are still no global standards or consensus on regulating these new technologies, especially in AI systems that can capture subjects' non-conscious data (McStay, 2020).

There are already troubling tech companies' reports in more industrialized countries selling bias prone facial recognition algorithms to developing countries' governments and the private sectors, where regulation is more relaxed (Roussi, 2020). Moreover, cultural norms and conventions arise when vendors try to transplant their biometric datasets and other analytic measurements to another culture. For example, US companies such as *HireVue*, are selling their video-interview analytic wares to Japanese companies and schools with claims to predict whether or not a candidate will make a successful future employee. Although the local distributor redesigns *HireVue*'s interview questions to align with Japanese society's norms and conventions, the biometric algorithms are created by Western programmers (Nakamura Toru, personal communication, January 15, 2021).

The fact that the culture is not static (Boyd et al., 2011), the globalization and the hyperconnectivity of our digital life foster faster and more complex cultural additivity that generate new values and sub-cultures (Vuong et al., 2018; Vuong, 2016). For example, the #BlackLivesMatter and #MeToo movements have already generated a wealth of new anti-racist and anti-sexist sentiments and vocabularies that are now spreading to the modern workplace (Hao, 2020). This should raise concerns about the sophistication of the current modus operandi of emotional AI technologies—the Ekman's *universal emotion hypothesis* (1985).

The meteoric rise of AI-driven micro-targeting, misinformation, and fake news has already clarified the potential pandora's box of unregulated smart technologies (Woolley & Howard, 2018). Much of the design and use of current artificial intelligence has exposed some of our natural stupidity (Vuong et al., 2019). This is evident in the glaring AI's social sciences deficit in multiple research areas, from data structuring to algorithmic designs (Sloane & Moss, 2019). Our study suggests three fundamental concerns for future job seekers who will be governed and assessed in either small or large ways by non-human resource management. The first is a privacy concern. The increased accuracy of emotion-sensing biometric technologies relies on a further blurring of personal/employee distinctions and tapping real-time unconscious data streams. The second is a concern for explainability. As emotional AI and its machine learning capabilities move toward greater complexity levels in automated thinking, many technologists believe that it will not be clear, even to the creators of these systems, how decisions are reached (Mitchell, 2019).

Finally, at a deeper biopolitical level, emotional AI represents an emerging era of automated governance where Foucauldian strategies and techniques of control are relegated to software systems in which "people willingly and voluntarily subscribe to and desire their logic, trading potential disciplinary effects against benefits gained" (Kitchin & Dodge, 2011, p. 11). Instead of physically monitoring and confining individuals in brick-and-mortar enclosures or enacting forms of control based on the body's exteriority, the *'algorithmic governmentality'* of emotion-sensing AI ultimately targets the mind and behavioral processes of workers to encourage their productivity and compliance (Mantello, 2016). Thus, with the emotional AI technologies becoming more pervasive in the business sectors, job-entry-level employees have significant challenges to retain privacy over their emotional lives and, by extension, their dignity, and autonomy. An individual's inability to negotiate the terms of such kinds of workplace power relations (Marciano, 2019) is compounded by the current absence of legal or ethical oversight.

In conclusion, the empirical cross-cultural and socio-demographic discrepancies observed in this paper seek to promote awareness and discussion and serve as a platform for further intercultural research on the ethical and social implications of emotional AI as a preeminent tool in non-human resource management.